\documentclass[12pt,preprint]{aastex}

\slugcomment{Submitted to the Astrophysical Journal}
\shorttitle{Large--Scale Intrinsic Alignment of Galaxy Images}
\shortauthors{Brainerd et al.}

\begin{document}

\title{Large--Scale Intrinsic Alignment of Galaxy Images}

\author{Tereasa G.\ Brainerd, Ingolfur Agustsson, Chad A.\ Madsen}
\affil{Boston University, Institute for Astrophysical Research, 725
Commonwealth Ave., Boston, MA 02215}
\email{brainerd@bu.edu, ingolfur@bu.edu, cmadsen@bu.edu}

\and
\author{Jeffrey A.\ Edmonds}
\affil{Boston University, Dept.\ of Religious and Theological Studies,
145 Bay State Road, Boston, MA 02215}
\email{jedmonds@bu.edu}

\begin{abstract}
We compute the two--point image correlation function for bright galaxies
in the seventh data release of the Sloan Digital Sky Survey (SDSS)
over angular scales $0.01' \le \theta \le 120'$ and projected separations
$0.01~{\rm Mpc} \le r_p \le 10~{\rm Mpc}$.  We restrict our analysis to SDSS
galaxies with accurate spectroscopic redshifts,
and we find strong
evidence for intrinsic alignment of the galaxy images.
On scales greater than 
$r_p \sim 40$~kpc, the intrinsic alignment of the SDSS galaxy images
compares well with the intrinsic alignment of galaxy images in a
$\Lambda$CDM universe,
provided we impose Gaussian--random errors on
the position angles of the theoretical galaxies with a dispersion of $\sigma_\phi =
25^\circ$.  Without the inclusion of these errors,
the amplitude of the two--point image correlation
function for the theoretical galaxies is a factor of $\sim 2$ higher than it is
for the SDSS galaxies.  We interpret this as a combination of modest position
angle errors for the SDSS galaxies, as well as a need for modest
misalignment of mass and light in the theoretical galaxies.
The intrinsic alignment of the SDSS galaxies shows no
dependence on the specific star formation rates of the galaxies and, at most,
a very weak dependence on the colors and stellar masses of the galaxies. At the
$\sim 3\sigma$ level, however, we find an indication that the images of
the most luminous SDSS galaxies are more strongly aligned with each other than
are the images of the least luminous SDSS galaxies.
\end{abstract}

\keywords{cosmology: observations --- galaxies: formation --- galaxies: halos ---
large-scale structure of the universe --- methods: statistical}

\section{Introduction}

The intrinsic alignment of galaxy images 
is of great interest for several reasons.  First, observations of
intrinsic alignments of galaxies over large scales constitute a test
of galaxy formation models.  Numerous numerical and analytical
studies have shown that galaxies should be intrinsically aligned with
each other over large scales in the universe (e.g., 
Heavens \& Peacock 1988;
Heavens et al.\ 2000;
Catelan et al.\ 2001; Crittenden et al.\ 2001, 2002; Croft \& Metzler 2001; 
Lee \& Pen 2001; Jing 2002; Mackey et al.\ 2002; Porciani et al.\ 2002;
Aubert et al.\ 2004; Heymans et al.\ 2004, 2006; Hirata \& Seljak 2004;
Lee et al.\ 2008; 
Okumura et al.\ 2009; Schneider \& Bridle 2009).  Observations of
intrinsic alignments of galaxies can, therefore, be used as consistency
checks on theories of galaxy formation.  Further, they can be used
to place constraints on the degree to which the images of luminous galaxies
are aligned with their dark matter halos.

In addition, intrinsic alignments of galaxies could be an important
contributor of ``false signal'' in studies that seek to measure weak
gravitational lensing by large-scale structure (a.k.a.\ cosmic shear).
To a first approximation, weak lensing by large-scale structure tends to
make galaxies that are nearby to each other on the sky ``point'' in the same
direction (see, e.g., Fig.\ 1 of Blandford et al.\ 1991).  Cosmic shear
requires substantial depths (i.e., moderate to large source 
redshifts) in order to be detectable.  Even still, in a deep, weak lensing
data set
care must be taken to account for any
contribution of intrinsic galaxy alignments to the observed
signal and, hence, in the subsequent quantification of cosmic shear.  
In principle this can be done either by eliminating
galaxies that are nearby to one another (in both position on the sky and 
redshift) or by modelling the expected contributions of intrinsic alignments
to the measured signal (see, e.g., 
Heymans \& Heavens 2003; King \& Schneider 2003;
Heymans et al.\ 2004; Takada \& White 2004;
Bridle \& King 2007; Joachimi \& Schneider 2008; Kitching et al.\ 2008; Zhang 2008).
Therefore, in order to obtain the best constraints on weak lensing by large-scale
structure, it is important to know the degree to which galaxies are intrinsically
aligned, including the scales over which the intrinsic alignment manifests
in our universe.

The first tentative detections of intrinsic alignments of galaxy images
(Brown et al.\ 2003; Heymans et al.\ 2004) did not include spectroscopic redshift
information that could be used to identify galaxies that are physically nearby
to one another.  More recent detections (Agustsson \& Brainerd 2006a; Lee \& Pen
2007; Okumura et al.\ 2009) have used both photometric and spectroscopic 
information from the Sloan Digital Sky Survey (SDSS; 
Fukugita et al.\ 1996; York et al.\ 2000;
Hogg et al.\ 2001; Smith
et al.\ 2002; Strauss et al.\ 2002; Ivezi\'{c} et al.\ 2004)) 
in order to restrict
the analysis to galaxies that are nearby to one another and, hence, to 
optimise the probability of detecting the signature of intrinsic alignment.

Agustsson \& Brainerd (2006a; hereafter AB06a) used the fourth data release (DR4)
of the SDSS to investigate the degree to which the images of satellite galaxies 
were aligned with their ``host'' galaxies.  The sample consisted of
3180 host galaxies that were relatively isolated in space (more isolated than
the Milky Way or M31) and 4289 satellite galaxies.  For projected separations
less than $r_p \sim 30$~kpc, AB06a found a tendency for the images of the 
satellite galaxies to be aligned in the same direction as the images of their
hosts.  Lee \& Penn (2007; hereafter LP07) used 434,849 galaxies from the
sixth data release of the SDSS to investigate differences between
the intrinsic alignments of red and blue galaxies as a function of 
three-dimensional separation.  LP07 restricted their analysis to galaxies
that had spectroscopic redshifts and $M_r \le -19.2$.  Over separations
$r \le 5h^{-1}$~Mpc, LP07 found a tendency for the images of the
red galaxies to be aligned.  LP07 also found a tendency for the images of
the blue galaxies to be aligned, but only over separations $r \le 3 h^{-1}$~Mpc.

Perhaps the most convincing detection of intrinsic alignment of galaxy images
to date is
the work of Okumura et al.\ (2009; hereafter OJL).  OJL restricted
their analysis to the 83,773 Luminous Red Galaxies (LRGs) in the SDSS and
computed the two--point image correlation function over three dimensional
separations $r \sim 1h^{-1}$~Mpc to $r \sim 130h^{-1}$~Mpc.  
OJL found a strong detection of the two--point image correlation function
for the images of the SDSS LRGs over these scales.  However, when OJL 
compared their results to a prediction of the intrinsic alignment expected for
elliptical galaxies in a $\Lambda$CDM universe, 
they found that the predicted two--point
image correlation function had an amplitude that was a factor $\sim 4$ larger
than they observed for the SDSS LRGs.  In order to compute the theoretical
image correlations, OJL made the common assumption that luminous elliptical
galaxies share the shapes of their dark matter halos (i.e., perfect alignment
of mass and light).  OJL then interpreted the discrepancy between the observed and
predicted two--point image correlation functions as 
evidence for a substantial misalignment of mass and light in elliptical galaxies
($\sigma_\theta = 35.4^{+4.0}_{-3.3}$~degrees).

Given the observed anisotropic distribution of satellite galaxies
around relatively isolated host galaxies (e.g., Brainerd 2005; 
Agustsson \& Brainerd 2007; Azzaro et al.\ 2007; Bailin et al.\ 2008;
Siverd et al.\ 2009),
as well as observations of weak gravitational lensing by the
flattened halos of field galaxies
(e.g., Hoekstra et al.\ 2004; Mandelbaum et al.\ 2006;
Parker et al.\ 2007), it would be surprising if the images of elliptical
galaxies were as badly misaligned from their halos as is suggested by OJL.
In the case of satellite galaxies, the satellites of relatively
isolated red host galaxies are located preferentially nearby the major axes
of their hosts, both in the SDSS and in $\Lambda$CDM universes
(e.g., Agustsson \& Brainerd 2007).  If 
the images of the red $\Lambda$CDM host
galaxies are misaligned with respect to their dark matter
halos by a very large amount, the anisotropy in the locations
of the satellites is reduced well below the level that is detected for
the satellites of red SDSS host galaxies.

In the case of weak lensing by flattened halos, one essentially
has to assume a reasonable alignment of mass and light in the lenses in order
to search for an anisotropy in the weak lensing signal.  That is, if mass and
light are badly misaligned for the lens galaxies, one will measure an isotropic
lensing signal on average since the only symmetry axes that one can use for
the measurement are the symmetry axes of the lensing galaxies themselves.  
Although detection
of anisotropic galaxy--galaxy lensing is difficult, it has been
tentatively observed for red lens galaxies (e.g.,  
Mandelbaum et al.\ 2006; Parker et al.\
2007).
This is an indication, then, that mass and light are at least reasonably 
well--aligned in the case of red field galaxies.   Similarly, in their
study of strong gravitational lenses, Keeton et al.\
(1998) found that the mass and light of early--type galaxies was aligned
to within $\sim 10^\circ$.

Here we investigate the intrinsic alignment of galaxy images in the 
seventh data release (DR7; Abazajian et al.\ 2009) of the SDSS.  
We restrict our analysis
to galaxies that are intrinsically rather bright, have well-determined 
spectroscopic redshifts, and which are located at sufficiently low redshifts
that their image shape parameters should be determined reasonably well.
The paper is organized as follows.  In \S2 we describe the subset of the
SDSS DR7 galaxies that we use for our study, we compute the intrinsic
alignment of the SDSS galaxies using a two--point image correlation 
function, and we subject the data to a number of null tests in order to 
check for the presence of systematics.  In \S3 we compare the intrinsic
alignment of the images of SDSS galaxies to the expectations for galaxies
in a $\Lambda$CDM universe.  In \S4 we investigate the degree to which 
the intrinsic alignment signal depends upon various physical parameters:
color, luminosity, stellar mass, and star formation rate.  Lastly, in \S5
we present a discussion of our results and our conclusions.
Throughout we adopt cosmological parameters 
$H_0 = 73$~km~sec$^{-1}$~Mpc$^{-1}$, $\Omega_{m0} = 0.25$, and
$\Omega_{\Lambda 0} =
0.75$.

\section{Intrinsic Alignments of SDSS Galaxy Images}

For our observational dataset we select all SDSS DR7 galaxies
with Petrosian apparent magnitudes
in the range $13.0 \le r \le 17.77$, absolute magnitudes in the range
$-23.0 \le M_r + 5 \log_{10} h \le -19.0$, and spectroscopic redshifts
with $\rm{zconf} > 0.9$.  There are a total of 530,397 galaxies in 
the SDSS DR7 that satisfy these criteria.  The median redshift of the
sample is $z_{\rm med} = 0.09$, and the maximum redshift is 
$z_{\rm max} = 0.20$.
In analogy to weak 
lensing analyses, we define a complex image shape parameter for
each of the galaxies as 
\begin{equation}
\vec{\chi} \equiv \epsilon e^{2i\phi},
\end{equation}
where $\epsilon = (a-b)/(a+b)$, $a$ is the semi-major axis of the 
galaxy, $b$ is the semi-minor axis of the galaxy, and $\phi$ is its
position angle.  Throughout we use shape parameters that have been 
computed using the SDSS $r$-band photometry.  

Also in analogy with the statistical measures that are commonly used to quantify
weak lensing by large--scale structure, we
define a two--point correlation function for galaxy images as
\begin{equation}
C_{\chi\chi} (\theta) \equiv 
\left< \vec{\chi}_{1} \cdot
\vec{\chi}_{2}^\ast \right>_{\theta}
\end{equation}
which measures the degree to which the image of galaxy 1 is aligned
with the image of galaxy 2, averaged over angular scales $\theta \pm d\theta/2$
(e.g., Blandford et al.\ 1991).  The function $C_{\chi\chi}(\theta)$
is positive if the major axes of the galaxies are aligned with
one another on average, it is
zero if the major axes of the galaxies are oriented randomly on average, and it
is negative if the major axes are oriented at right angles to 
one another on average.

Since we expect that galaxies will be not be intrinsically aligned over
extremely large 
physical scales in the universe, we restrict our analysis to pairs of
galaxies that are relatively close to each other in velocity space.  
That is, pairs of galaxies that are separated by many hundreds of megaparsecs 
along the line of sight are unlikely to be intrinsically aligned,
while pairs of galaxies that are separated by a few tens of megaparsecs along the
line of sight should be intrinsically aligned to at least  
some degree.
This in mind, we compute $C_{\chi\chi}(\theta)$ for pairs of galaxies with
line of sight velocity differences $|dv| \le 5000$~km~sec$^{-1}$.  Fig.\ 1
shows the result of this computation for SDSS DR7 galaxies that
are separated by
angles $0.01' \le \theta \le 120'$ on the sky.  The function has been 
computed using angular bins that are equally--spaced in the logarithm, 
and the error bars show the
standard deviation in the mean of $C_{\chi\chi}(\theta)$.

\begin{figure}
\centerline{\scalebox{0.70}{\includegraphics{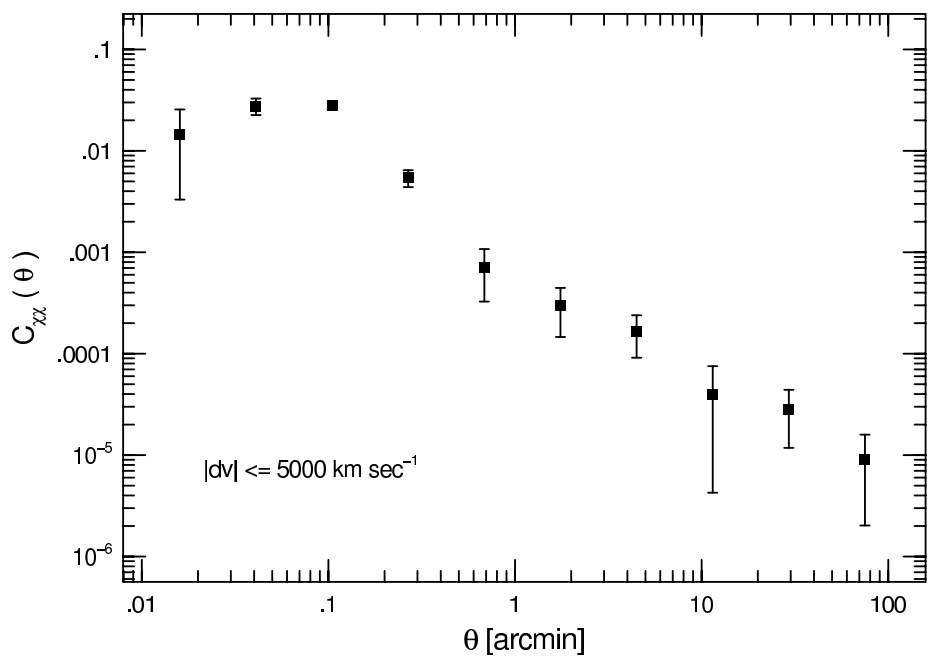} } }%
\vskip -0.0cm%
\caption{Two--point correlation function of image shapes for
SDSS galaxies computed using pairs of galaxies with angular
separations $0.01' \le \theta \le 120'$.  
Pairs of galaxies are restricted to line of sight velocity 
differences $|dv| \le 5000$~km~sec$^{-1}$.   Error bars show
the standard deviation in the mean of $C_{\chi\chi}(\theta)$.}
\label{fig1}
\end{figure}

\begin{figure}
\centerline{\scalebox{0.70}{\includegraphics{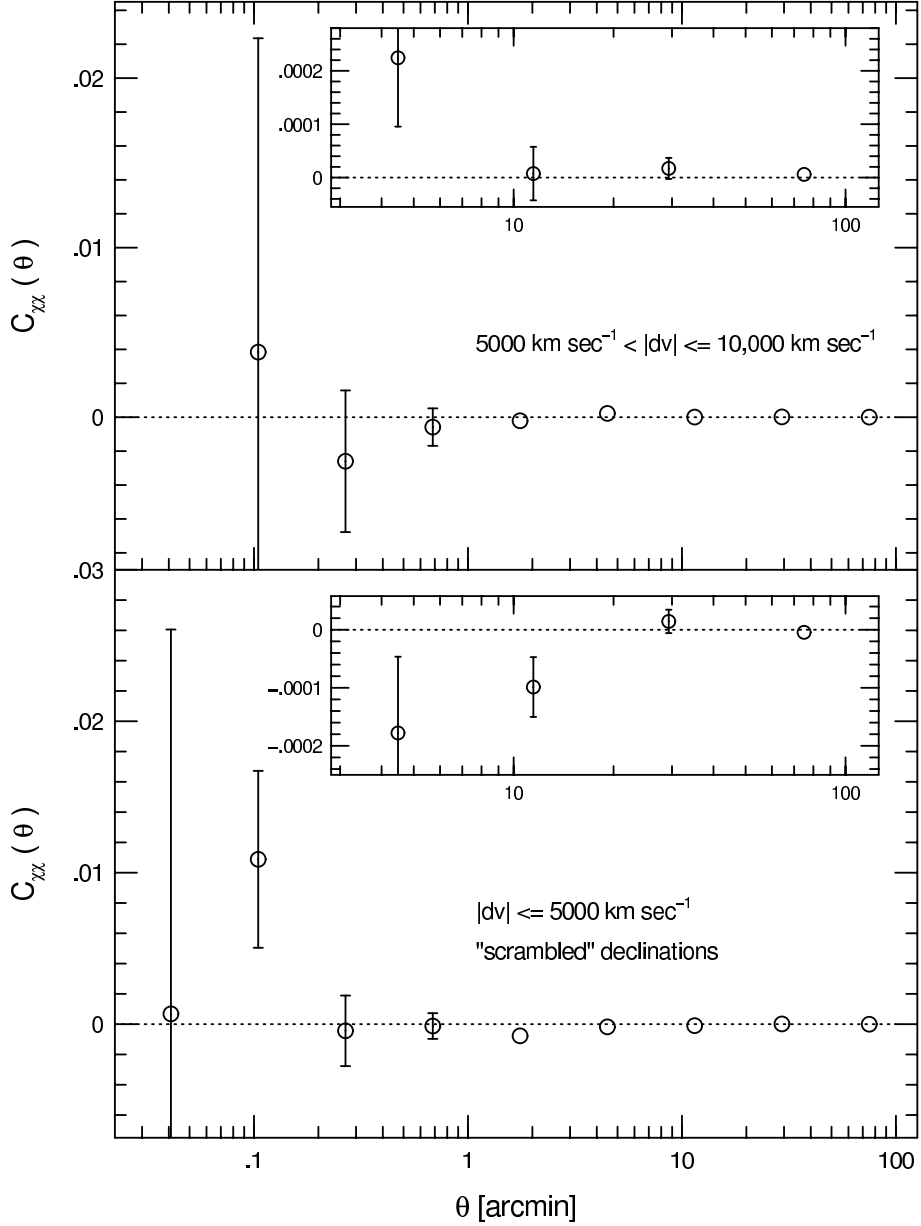} } }%
\vskip -0.0cm%
\caption{Null tests using the two--point
correlation function of image shapes for SDSS galaxies.  Top
panel: $C_{\chi\chi}(\theta)$ computed as in Fig.\ 1, but here
the pairs of galaxies are separated by line of sight
velocity differences of 5000~km~sec$^{-1}$ $< |dv| \le$
10,000~km~sec$^{-1}$.  Bottom panel: $C_{\chi\chi}(\theta)$ computed
as in Fig.\ 1, but here the declinations of the galaxies have been
scrambled.  Error bars are omitted in the case that the error bar is
smaller than the data point.  Insets show $C_{\chi\chi}(\theta)$ for the
last 4 bins on an expanded scale. Note that in the top panel there are
no pairs of galaxies that fall into the first two angle bins of Fig.\ 1.
In the bottom panel, there are no pairs of galaxies that fall into the first
angle bin of Fig.\ 1.
}
\label{fig2}
\end{figure}

\begin{figure}
\centerline{\scalebox{0.70}{\includegraphics{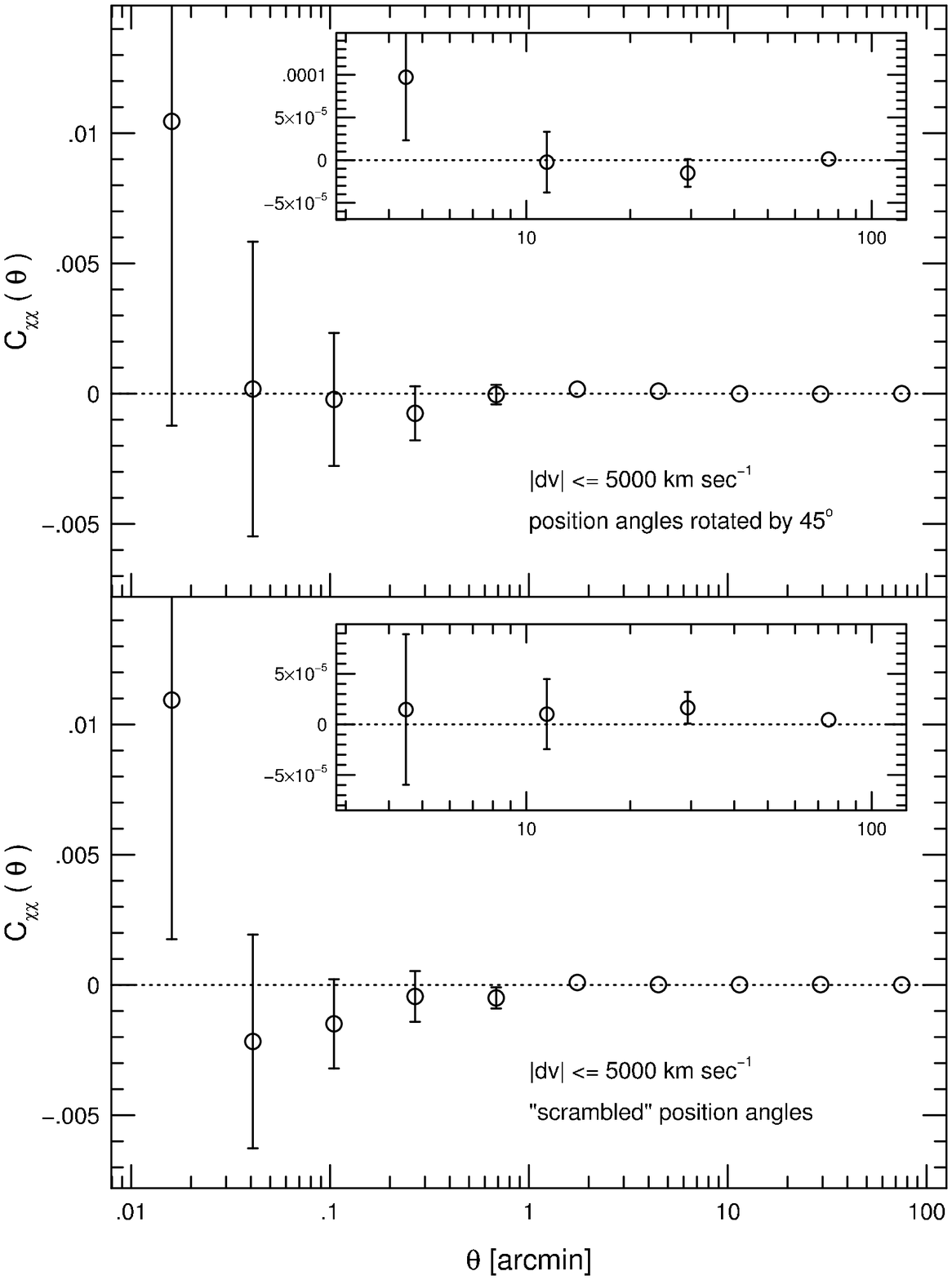} } }%
\vskip -0.0cm%
\caption{Null tests using the two--point
correlation function of image shapes for SDSS galaxies.  Top
panel: $C_{\chi\chi}(\theta)$ computed as in Fig.\ 1, but here
the position angles of the galaxies have been rotated by $45^\circ$.
Bottom panel: $C_{\chi\chi}(\theta)$ computed
as in Fig.\ 1, but here the position angles of the galaxies
has been scrambled.
Error bars are omitted in the case that the error bar is
smaller than the data point.  Insets show $C_{\chi\chi}(\theta)$ for the
last 4 bins on an expanded scale. 
}
\label{fig3}
\end{figure}

It is clear from Fig.\ 1 that the images of the SDSS galaxies have
a small but detectable tendency to be aligned with each other over 
large angular scales on the sky.  To investigate whether the result
in Fig.\ 1 is genuine (and, therefore, indicative of intrinsic
alignment of the galaxies) or whether it
can be attributed to a systematic, we perform
a number of null tests.  First, we compute $C_{\chi\chi}(\theta)$
exactly as we did in Fig.\ 1, but we restrict the analysis to 
pairs of galaxies that are separated by large amounts in
velocity space (5000~km~sec$^{-1}$ $< |dv| \le$ 10,000~km~sec$^{-1}$).
Second, we randomize the declinations of the galaxies (i.e., 
swapping the declinations amongst the galaxies so that galaxy $i$
is assigned the declination of galaxy $j$) while maintaining
the observed right ascensions and redshifts of the galaxies.  Third,
we perform a classic weak lensing null test
in which we rotate the observed position angles by $45^\circ$.
Lastly, we randomize the position angles of the 
galaxies by swapping the observed position angles amongst the
galaxies (i.e., the position angle of galaxy $i$ is assigned
to galaxy $j$).  In the last three null 
tests we restrict the pairs of galaxies to be separated by $|dv| \le
5000$~km~sec$^{-1}$, as we did for the calculation in Fig.\ 1.  The
function $C_{\chi\chi}(\theta)$ ought to vanish for all of the
null tests if the signal shown in Fig.\ 1 is genuine and not due to
an artifact in the data.

The results of the first two null tests are shown in Fig.\ 2 and the
results of the second two null tests are shown in Fig.\ 3.  In all cases,
$C_{\chi\chi}(\theta)$ for the null tests
is consistent with the images of the galaxies
being uncorrelated.  Therefore, the alignment exhibited in Fig.\ 1 is
unlikely to be caused by a systematic.

\section{Comparison to Intrinsic Alignment of Galaxy Images in a $\Lambda$CDM
Universe}

We now investigate whether the apparent intrinsic alignment of the SDSS
galaxies in the previous section
is consistent with the expectations for bright galaxies in 
a $\Lambda$CDM universe.  To do this we make use of the Millennium Run Simulation
(MRS; Springel et al.\ 2005) and the corresponding MRS mock galaxy redshift catalog
that was constructed by Blaizot et al.\ (2005) using the Mock Map Facility (MoMaF). 
The MRS is a large simulation of the growth of structure in
a $\Lambda$CDM universe that has both high spatial resolution and high mass 
resolution (softening length of $5h^{-1}$~kpc and particle mass
$m_p = 8.6\times 10^8 h^{-1}~M_\odot$).  Although the MRS particle files are not
publicly-available due to their enormous size, information about the
dark matter halos, subhalos, and
the luminous galaxies (as determined from 
semi-analytic galaxy formation codes) is 
publicly-available\footnote{http://www.mpa-garching.mpg.de/millennium/}.  The
mock redshift catalog for the MRS galaxies is also 
publicly-available\footnote{http://galics.iap.fr}.

Stellar masses, star formation rates, bulge-to-disk ratios,
luminosities, colors, and dark matter halo masses
are all publicly-available for the MRS galaxies.  Importantly
for our work, however, is that there is no information on the observed
{\it shapes} of the luminous galaxies in the MRS.  In order to compare the intrinsic
alignment of the SDSS galaxies to the expectations of a $\Lambda$CDM universe,
we therefore need to choose a method by which to {\it assign} shapes to the
MRS galaxies.  To do this, we use the B--band bulge-to-disk ratios
($\Delta M(B) = M(B)_{\rm bulge}-M(B)_{\rm total}$) from 
the semi-analytic galaxy formation model of De Lucia et al.\ (2006) to 
classify the MRS galaxies into three broad categories: ellipticals,
lenticulars, and spirals.  Following De Lucia et al.\ (2006), we consider
MRS galaxies with $\Delta M(B) < 0.4$ to be ellipticals and those with
$0.4 \le \Delta M(B) \leq 1.56$ to be lenticulars.  In addition, we assume that
all MRS galaxies with $\Delta M(B) > 1.56$ that will be used in our analysis
are spiral galaxies.  In principle galaxies  with $\Delta M(B) > 1.56$ could
be irregulars, but given that we are restricting our
analysis to galaxies with rather bright luminosities
($-23.0 \le M_r + 5 \log_{10} h \le -19.0$; see \S2), it is
unlikely that any MRS galaxy with $\Delta M(B) > 1.56$ that meets our minimum
luminosity criterion would be an irregular.

In order to assign observed shapes to the MRS galaxies (i.e., observed position
angles and ellipticities), we adopt a very simple prescription.  We assume
that ellipticals share the same shapes as their dark matter halos, and we
assume that all disk galaxies (i.e., the spirals and lenticulars) are circular
disks whose angular momentum vectors are perfectly aligned with the net angular 
momentum vectors of their dark matter halos.   The angular momentum vectors of the 
dark matter halos that we require for the disk galaxies are publicly-available; 
however, the moments of inertia of the halos are not publicly-available.
Therefore, in order to assign shapes to the elliptical galaxies we need to
adopt a proxy for the mass distribution within the halos.  Previous work
using the $\Lambda$CDM GIF simulation
has shown that small satellite galaxies
trace the shapes of the halos that surround the large, primary galaxies 
about which the satellites are orbiting (Agustsson \& Brainerd 2006b).  Therefore,
to determine the shape of a halo surrounding an elliptical MRS galaxy,
we use the locations of its satellites as a proxy for the mass distribution
of the halo.  We restrict our analysis to elliptical galaxies that have at 
least 4 satellites within the virial radius, and using these we compute a
3-dimensional ellipsoid that defines the shape of the galaxy. 

Of order 5\% of the MRS galaxies that fall within
the luminosity range that we have adopted have been stripped of all of
their dark matter due to numerical effects.  These are referred to as ``type 2''
galaxies in the De Lucia \& Blaizot (2007) semi-analytic model for the
MRS galaxies, and all
are satellite galaxies that are orbiting within the dark matter halo of
a larger galaxy.  Since there is no dark matter substructure around the
type 2
galaxies, we cannot assign observed shapes to them and we therefore do not use them
in our analysis.  The loss of these galaxies will restrict our ability
to measure the intrinsic alignments on small physical scales (i.e., scales
much less than the virial radii of the halos of large galaxies) since these
galaxies contribute a great deal to the pair counts at small separations.
On larger scales, however, the pair counts are heavily dominated by the 
galaxies for which we can measure the shapes, and the loss of the type 2 
galaxies from the analysis will have no affect on the results.  In total, 
there are $3.1\times 10^6$ MRS galaxies that fall within our adopted
luminosity range and for which we can assign shapes. 

To obtain the observed position angles and ellipticities of the MRS galaxies, we 
project the 3-dimensional shapes of the galaxies (i.e., ellipsoids or circular disks)
onto the sky using the galaxy locations from the mock redshift
survey of Blaizot et al.\ (2005).  We then compute the complex image shape 
parameter, $\vec{\chi}$, for each MRS galaxy in exactly the same way as it
was computed for the SDSS galaxies. 

To compare the degree of alignment of the galaxy images in the SDSS DR7
and in the MRS, we compute the two--point image correlation function as a function
of projected separation on the sky, $C_{\chi\chi}(r_p)$.  We expect this to 
provide a more reasonable comparison than the two--point image correlation function
computed as a function of angular separation, $C_{\chi\chi}(\theta)$. This is
because, although the redshift distributions of the 
SDSS galaxies and the MRS mock redshift survey galaxies are very similar, they
are not identical.  This will affect the number of pairs found
a given angular separation, $\theta$, somewhat more than it will affect the number
of pairs found to be separated by a given projected separation, $r_p$.  This
is due to the fact that when we calculate $C_{\chi\chi}(\theta)$ we do not
account for the specific value of the line of sight velocity difference 
between the two galaxies (i.e., we simply accept all pairs of galaxies with
$|dv| \le 5000$~km~sec$^{-1}$).  In the case of $C_{\chi\chi}(r_p)$ we make
use of the specific value of the
velocity difference in addition to the angular separation.

In both the SDSS and MRS we define the projected separation between a
pair of galaxies to be
\begin{equation}
r_p \equiv \frac{\theta}{2} \left(D_{A1} + D_{A2} \right) 
\end{equation}
where $\theta$ is the angular separation between the galaxies (in radians),
$D_{A1}$ is the angular diameter distance to galaxy 1, and $D_{A2}$ is
the angular diameter distance to galaxy 2.  We compute the angular
diameter distances for the SDSS and MRS galaxies in the standard way
using their observed 
redshifts and our adopted cosmological parameters.   The resulting two--point
correlation functions are shown in Fig.\ 4.

As in \S2, we restrict our analysis in Fig.\ 4 
to galaxies with absolute magnitudes
in the range $-23.0 \le M_r + 5 \log_{10} h \le -19.0$.
We then compute
$C_{\chi\chi}(r_p)$ for galaxy pairs separated by $0.01~{\rm Mpc} \le
r_p \le 10~{\rm Mpc}$. 
From Fig.\ 4, 
we see that the degree of intrinsic alignment of the galaxy images in the
SDSS is comparable to that in a $\Lambda$CDM universe.  In the first
bin, $C_{\chi\chi}(r_p)$ for the MRS galaxies sits quite low compared to 
$C_{\chi\chi}(r_p)$ for the SDSS galaxies.  The discrepancy is most likely
due to the fact that on such small scales (i.e., $r_p \sim 15$~kpc) we 
simply do not have the ability to measure $C_{\chi\chi}(r_p)$ accurately 
for the MRS galaxies.
Roughly 80\% of the galaxy pairs in the MRS that ought to go into the calculation
of $C_{\chi\chi}(r_p)$ in the first bin include at least one ``type
2'' galaxy (i.e., a galaxy for which we cannot assign a shape due to the 
fact that it has been stripped of all dark matter).  That is, on the
smallest physical scales the two--point correlation function of image shapes is
dominated by the degree to which large galaxies and their satellites are
aligned with each other,
and we simply do not have the ability to measure this well for the
MRS galaxies.  To further emphasize that the primary contributors to
intrinsic alignments on small
scales are the images of satellites and
the galaxies which they orbit, we also show in Fig.\ 4 (stars)
the small-scale intrinsic alignment found by 
AB06a for the images of relatively isolated SDSS host galaxies and
their satellites. 
The agreement between the results of AB06a and our
present analysis of SDSS DR7 galaxies is quite good, demonstrating that
intrinsic alignment of galaxy images at the smallest projected radii is 
largely attributable to galaxies that are physically quite close together
in space.

\begin{figure}
\centerline{\scalebox{0.70}{\includegraphics{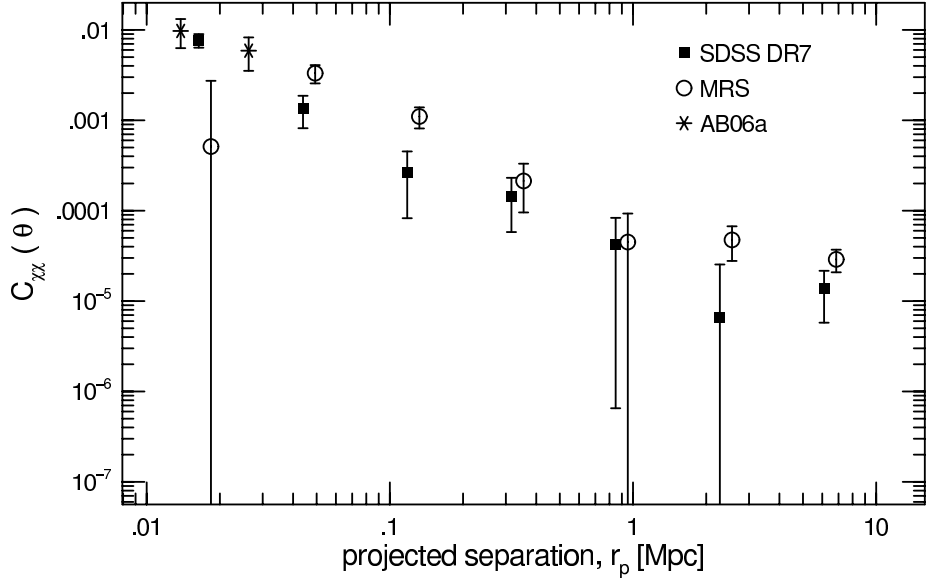} } }%
\vskip -0.0cm%
\caption{
Two--point correlation function of galaxy images in the SDSS DR7 (squares)
and the Millennium Run Simulation (circles) computed as a function of
projected separation on the sky.  
The data points have
been slightly offset in the horizontal direction for clarity.
Also shown for comparison (stars) is the small-scale intrinsic alignment found by
AB06a for the images of host and satellite
galaxies in the SDSS DR4.
}
\label{fig4}
\end{figure}

\begin{figure}
\centerline{\scalebox{0.70}{\includegraphics{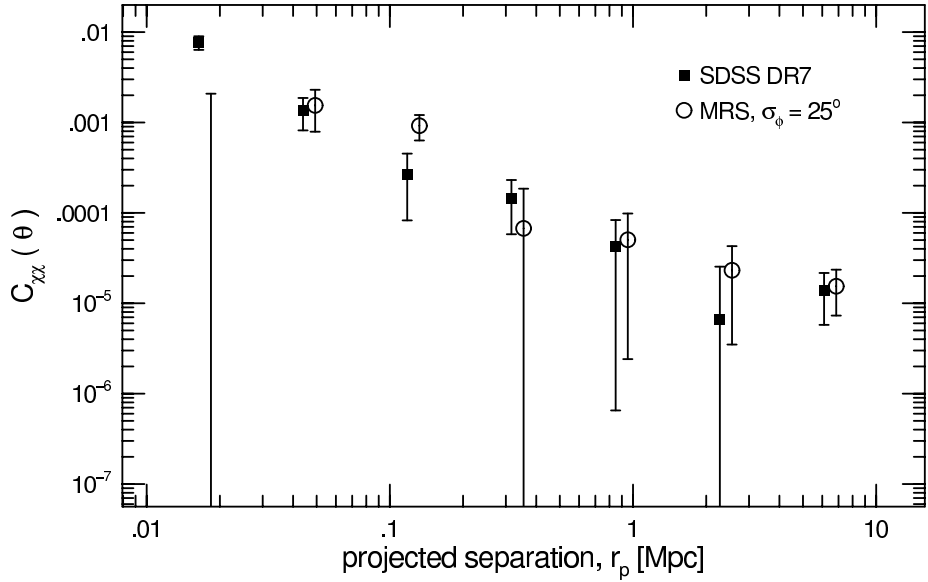} } }%
\vskip -0.0cm%
\caption{
Same as Fig.\ 4, except that Gaussian--random errors with zero mean
and dispersion $\sigma_{\phi} = 25^\circ$ have been added to the 
position angles of the MRS galaxies.  The data points have
been slightly offset in the horizontal direction for clarity.
}
\label{fig5}
\end{figure}

Fig.\ 4 shows that on scales larger than $r_p \sim 40$~kpc, the 
intrinsic alignment of the MRS galaxy images is somewhat stronger
than that of the SDSS galaxy images.
Overall, the amplitude differs by a factor of order 2 on these scales.
In their analysis of the intrinsic alignments of galaxies,
OJL also found that the images of elliptical
galaxies in a $\Lambda$CDM universe showed a greater degree of intrinsic
alignment than did the images of the SDSS luminous red galaxy (LRG) sample.
Like us, OJL assumed that elliptical galaxies shared the shapes
of their halos and they found that the two--point correlation function of image
shapes was a factor of order 4 larger for the $\Lambda$CDM ellipticals than it
was for the SDSS LRGs.  To explain this discrepancy, OJL
argue that the $\Lambda$CDM galaxy images must be misaligned with respect 
to their halos,
and that the misalignment is quite large indeed
($\sigma_\theta = 35.4^{+4.0}_{-3.3}$
degrees).

We now consider the degree to which the position angles of the MRS galaxy images
must differ from the values used in Fig.\ 4 in order to yield a two--point
image correlation function that matches that of the SDSS galaxies. 
To do this, we assign Gaussian--random errors to the position angles of the MRS
galaxies and we recompute $C_{\chi\chi}(r_p)$ using these ``corrupted'' position
angles.  Shown in Fig.\ 5 is the result 
for $C_{\chi\chi}(r_p)$ in which
Gaussian--random errors with zero mean and dispersion 
$\sigma_{\phi} = 25^\circ$ have been assigned to the position angles
of the MRS galaxies.   From this figure, then, it is clear that on 
scales larger than $r_p \sim 40$~kpc, there is good agreement between the
intrinsic alignments of the SDSS and MRS galaxies if the typical error in the
position angle is of order $25^\circ$.  This is substantially smaller than
the galaxy--halo misalignment required by OJL, but is
consistent with the galaxy--halo misalignment required by Faltenbacher
et al.\ (2009) in their study of the alignments of bright, red galaxies 
with local large--scale structure.  In addition, if we increase the dispersion
for the position angles of the MRS galaxies to $\sigma_\phi = 35^\circ$, 
$C_{\chi\chi}(r_p)$ for the MRS galaxies becomes consistent with zero (i.e., 
we can no longer detect intrinsic alignments using our theoretical galaxy
images if the error in the position angle has such a large dispersion).

We also note that the shape parameters of the images of the SDSS
galaxies cannot be measured with infinite accuracy; i.e., there is some 
noise in the determination of $\vec{\chi}$ for all SDSS galaxies.  The function
$C_{\chi\chi}(r_p)$ is not particularly sensitive to errors in the ellipticities of
the galaxies, but it is highly sensitive to errors in the position
angles of the galaxies.  Therefore, any difference between the intrinsic alignments
of theoretical galaxies (e.g., the MRS galaxies in our case) and observed
galaxies can, at least in part, be attributed to observational errors associated
with position angle measurements.  Unfortunately, position angle errors for 
the SDSS galaxies are not yet available in the archive.  Therefore, we cannot
place any direct constraints on what fraction of the error needed above is
due to observational errors (i.e., determination of the position angles) and
what fraction is due to misalignment of mass and light in galaxies.  If the
contributions are equal (and assuming Gaussian errors), we would require errors
of $\sim 17^{\circ}$ in both cases.  Such errors are not unreasonable given 
what few direct constraints we have on the alignment of mass and light in
galaxies, as well
as the fact that many SDSS galaxies will have position angles that are challenging
to determine (e.g., very round galaxies with smooth light profiles, or disk 
galaxies that have bright pockets of star formation and/or resolved spiral arms).

\section{Dependence of Intrinsic Alignments on Galaxy Properties}

\begin{figure}
\centerline{\scalebox{0.70}{\includegraphics{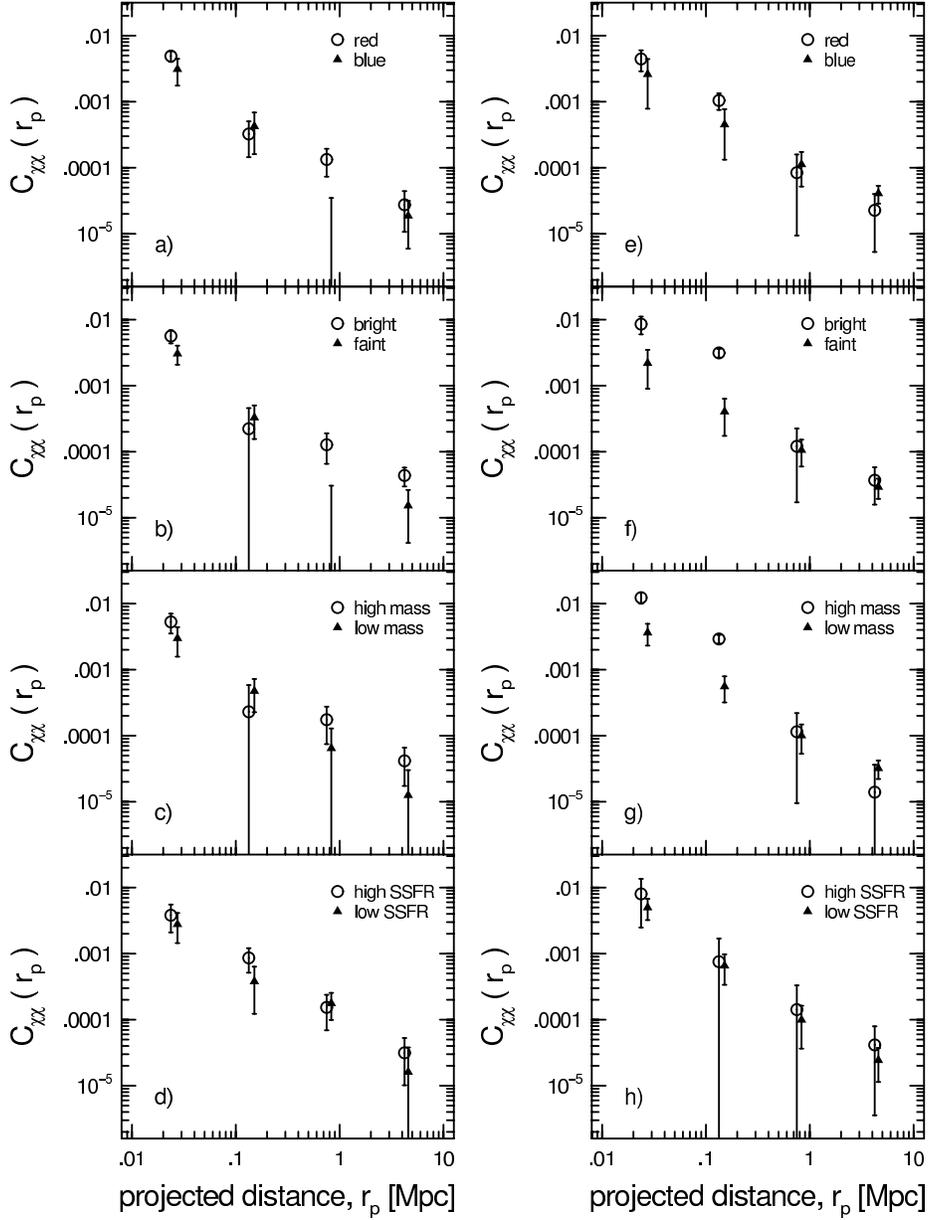} } }%
\vskip -0.0cm%
\caption{Dependence of the
two--point image correlation function for SDSS galaxies (left panels)
and MRS galaxies (right panels) on a number of physical properties.    
See text for details regarding the specific subdivisions of the galaxy
samples.
The data points have
been slightly offset in the horizontal direction for clarity.
}
\label{fig6}
\end{figure}

Next, we investigate the degree to which the intrinsic alignments of
the SDSS and MRS galaxies depend upon various physical properties:
$(g-r)$, luminosity, stellar mass, and star formation
rate.  While each of these properties is known for all of the MRS galaxies,
the stellar masses and star formation rates have only been
computed for SDSS galaxies that are contained within the DR4.
The stellar masses for the SDSS DR4 galaxies were computed by Kauffmann
et al. (2003) and the star formation rates were computed
by Brinchmann et al.\ (2004).  Our sample of SDSS galaxies
with measured stellar masses is therefore reduced to 291,728 and our sample of SDSS
galaxies with measured star formation rates is reduced to 297,284.

To divide our sample of galaxies into ``red'' and ``blue'' galaxies,
we first K--correct the SDSS galaxy colors to the present epoch using
the code by Blanton et al.\ (2003; v4\_1\_4).  We then fit the distributions
of $(g-r)$ colors of the SDSS galaxies by the sum of two Gaussians
(e.g., Strateva et al.\ 2001; Weinmann et al.\ 2006) and we find that
the division between the two Gaussians lies at $(g-r)=0.7$.  We therefore
define galaxies with $(g-r) < 0.7$ to be ``blue'' and galaxies with
$(g-r) \ge 0.7$ to be ``red''.  Imposing this criterion, we have
235,834 red galaxies in our sample and 294,563 blue galaxies.
To divide our sample of galaxies into a ``bright'' and ``faint'' half,
we simply use the median $r$-band absolute magnitude of the SDSS galaxies,
$M_r = -21.16$.  To divide the sample into a ``high stellar mass'' and 
``low stellar mass'' sample we use the median stellar mass of the SDSS DR4 galaxies:
$\log_{10}(M_\ast/M_\odot) = 10.61$.  Lastly, we define the specific star
formation rate (SSFR) to be the ratio of the star formation rate
(in $M_\odot$~yr$^{-1}$) to
the stellar mass (in solar units), and we divide the sample into a ``high SSFR''
and ``low SSFR'' sample using the median SSFR for the SDSS DR4 galaxies:
$\log_{10}({\rm SSFR / yr^{-1}}) = -9.91$.  

The results for the dependence of
the two--point image correlation function on $(g-r)$, luminosity, stellar
mass, and SSFR are shown in Fig.\ 6.  The left panels show
the results for the SDSS galaxies, and the right panels show the results
for the MRS galaxies.  Panels a) and e) show the dependence of 
$C_{\chi\chi}(r_p)$ on $(g-r)$, panels b) and f) show the dependence of
$C_{\chi\chi}(r_p)$ on luminosity, panels c) and g) show the dependence of
$C_{\chi\chi}(r_p)$ on stellar mass, and panels d) and h) show the dependence of
$C_{\chi\chi}(r_p)$ on the specific star formation rate (SSFR).  For the most
part, the error bars in Fig.\ 6 for the various subdivisions of the data
overlap each other, indicating a rather weak dependence of $C_{\chi\chi}(r_p)$
on the properties of the galaxies.  To supplement Fig.\ 6 we therefore compute an
average of $C_{\chi\chi}(r_p)$ over all values of the projected
separation that we have considered, $0.01~{\rm Mpc} \le r_p \le 10~{\rm Mpc}$,
and we show the results in Table~1.

In the case of the MRS galaxies, the value of $C_{\chi\chi}(r_p)$ averaged over
$0.01~{\rm Mpc} \le r_p \le 10~{\rm Mpc}$ shows essentially no dependence
on the properties of the galaxies.  That is, for all four subdivisions of
the data, the separate values of $\left< C_{\chi\chi} \right>$ agree to within
1$\sigma$.  In the case of the SDSS galaxies, there is no apparent 
dependence of $\left< C_{\chi\chi} \right>$ on SSFR.  There is a very weak
indication (at the 1.4$\sigma$ and 1.6$\sigma$ level, respectively) that
the red SDSS galaxies are more aligned than are the blue SDSS galaxies, and
that the high stellar mass SDSS galaxies are more aligned than are the low
stellar mass SDSS galaxies.  More convincing is the difference in the intrinsic
alignments of the high luminosity SDSS galaxies compared to the low luminosity
SDSS galaxies; at the 2.8$\sigma$ level the high luminosity SDSS galaxies
are more strongly aligned on average than are the low luminosity SDSS galaxies.
While not definitive, this last result is at least intriguing.

Finally, it is clear from the right hand panels of Fig.\ 6 that 
the dependence of $C_{\chi\chi}(r_p)$ on projected
separation is rather different for the bright and faint MRS galaxies (panel f), as well
as for the high and low stellar mass MRS galaxies (panel g).  For both
the faint and the low stellar mass MRS galaxies, $C_{\chi\chi}(r_p)$
appears to be a power law.  For both the bright and the
high stellar mass MRS galaxies, $C_{\chi\chi}(r_p)$ appears to
be a broken power law, with
a clear change in amplitude at projected radii of order 300~kpc.  On scales less
than about 300~kpc, $C_{\chi\chi}(r_p)$ for the high luminosity MRS galaxies 
is a factor of order 6 larger than $C_{\chi\chi}(r_p)$ for the low luminosity
MRS galaxies.  Similarly, on scales less than about 300~kpc, $C_{\chi\chi}(r_p)$
for the MRS galaxies with high stellar masses is a factor of order 4 larger
than $C_{\chi\chi}(r_p)$ for MRS galaxies with low stellar masses.  It is
hard to assess the significance of this, but it may be an indicator that in
a redshift survey that is significantly larger than the SDSS, 
we should expect to see substantial differences in the intrinsic alignments
of high and low luminosity galaxies, as well as high stellar mass and low
stellar mass galaxies, at least on scales less than about 300~kpc.

\section{Discussion and Conclusions}

We have computed the two--point image correlation function for bright
galaxies in the SDSS DR7.  The function withstands several null tests and, 
therefore, is more likely to be indicative of intrinsic alignments of the 
SDSS galaxy images rather than a systematic in the data. On scales greater
than $r_p \sim 40$~kpc, the observed two--point image correlation 
function for the SDSS galaxies has an amplitude that is a factor of
order two lower than our prediction for the intrinsic alignment of galaxy
images in a $\Lambda$CDM universe.  The discrepancy between the observed
and predicted functions can be eliminated by including a Gaussian--random
error on the position angles of the theoretical galaxy images with 
dispersion $\sigma_\phi = 25^\circ$.  

Previous studies of intrinsic alignments of galaxy images have also
concluded that the observed alignment of galaxy images is less than predicted.
In particular, OJL found that the observed
two--point image correlation function for the SDSS LRGs has an amplitude
that is a factor of order 4 lower than the predicted function.  To explain
the discrepancy, OJL claim that mass and light in large
elliptical galaxies must be misaligned by $\sigma_\theta = 35.4^{+4.0}_{-3.3}$
degrees.  Given the observed alignment of the satellite distribution around
large, red SDSS galaxies, as well as observations of weak and strong lensing
by early-type galaxies, we expect that such a large misalignment of mass and
light in the SDSS LRGs is unlikely.  Also, given the fact that the two--point
image correlation function is highly sensitive to errors in the measured 
position angles of the galaxies, at least some of the disagreement between
the observed and theoretical functions can be attributed to the finite accuracy
with which the position angles of the observed galaxies can be measured.  We 
therefore conclude that, at least in the case of the SDSS galaxies that we
have studied, the difference between the observed and predicted two--point
image correlation functions can be explained by a combination of modest
misalignment of luminous galaxies and their halos, as well as modest 
position angle errors for the observed galaxy images.

In their study of the intrinsic alignments of galaxy images in the 
sixth data release of the SDSS, LP07 concluded that for
separations greater than $3h^{-1}$~Mpc, the images of
red galaxies showed intrinsic
alignment, while the images of blue galaxies did not.  In our analysis
we find at best only weak evidence for a color dependence of 
the intrinsic alignments, with the red galaxies showing very slightly
more intrinsic alignment than the blue galaxies.
It is difficult to know why we obtain results
that are so different from LP07, but it
may have to do with a combination of
the redshifts over which the galaxies were selected
as well as the way in which red galaxies were separated from blue galaxies.
In LP07, the galaxies were divided into ``red'' and ``blue'' samples using
$(u-r)$ colors that were not K-corrected.  Only 20\% of LP07's galaxies
were ultimately classified as being blue, while 56\% of our galaxies are classified
as being blue.  In addition, LP07 used galaxies with redshifts as large
as $z=0.4$, while our sample is restricted to lower redshifts (maximum
redshift of $z=0.2$).  Generally, the redder is a galaxy, the more accurate
is a measurement of its shape (i.e., since red galaxies tend to have very
smooth light profiles).  It is also true that, in general,
the closer is a galaxy in redshift
space (for fixed luminosity), the more accurate is a measurement of its shape.
It is, therefore, possible that LP07 were able to measure an
intrinsic correlation of the red galaxy images but not the
blue galaxy images on scales $r > 3h^{-1}$~Mpc because: [i] the image shapes
of the red galaxies were far less noisy than those of the blue galaxies, and
[ii] there were simply too few blue galaxies in their sample to detect the 
intrinsic correlations in the presence of noisy image shapes (i.e., assuming
Poisson errors).

In addition to weak evidence for a color dependence of the intrinsic
alignments of galaxy images, we also find weak evidence that the degree
of intrinsic alignment for the SDSS galaxies is dependent upon the 
stellar masses (i.e., more massive galaxies show very slightly more 
alignment than do less massive galaxies).   More convincing is the 
dependence we find on luminosity for the SDSS galaxies; at the $\sim 3\sigma$
level the most luminous SDSS galaxies in our sample
show stronger intrinsic alignment
than do the least luminous galaxies. 

Lastly, we note that we only detect intrinsic alignment of galaxy images when
the galaxies are separated by less than 5000~km~sec$^{-1}$ along the line
of sight.  This bodes well for future weak lensing ``tomography'' experiments
that wish to eliminate pairs of galaxy images that may be intrinsically 
aligned before computing the weak lensing signal.  Assuming that the
large-scale intrinsic alignments of galaxy images do not persist far beyond
$|dv| \sim 5000$~km~sec$^{-1}$, this suggests that the use of
photometric redshifts with relatively modest accuracy should suffice to 
eliminate the intrinsic alignment signal from the sought-after weak lensing
signal.

\section*{Acknowledgments}

Support under NSF contract AST-0708468 (TGB, IA, CAM) is
gratefully acknowledged.   We are also deeply grateful for the public
release of the Millennium Run Simulation and its numerous associated
databases.
Funding for the SDSS and SDSS-II has been provided by
the Alfred P. Sloan Foundation, the Participating Institutions, 
the U.S. Department of Energy, the
National Science Foundation, the National Aeronautics and Space Administration,
the NSF, the Japanese Monbukagakusho, 
the Max Planck Society, and the Higher Education Funding Council for England.  
The SDSS is managed by the Astrophysical
Research Consortium for the Participating Institutions: the American
Museum of Natural History, Astrophysical Institute Postdam, University
of Basel, University of Cambridge, Case Western University, University
of Chicago, Drexel University, Fermilab, the Institute for Advanced Study, the Japan 
Participation Group, The Johns Hopkins University, the Joint
Institute for Nuclear Physics, the Kavli Institute for
Particle Astrophysics and Cosmology, the Korean Scientist Group, the
Chinese Academy of Sciences (LAMOST), Los Alamos National
Laboratory, the Max-Planck-Institute for Astronomy (MPIA), the
Max-Planck-Institute for Astrophysics (MPA), New Mexico State University, The
Ohio State University,
University of Pittsburgh, University of Portsmouth,
Princeton University, the U.S. Naval Observatory,
and the University of Washington.  The SDSS Web site is 
\url{http://www.sdss.org}.

\clearpage
\centerline{Table 1: $C_{\chi\chi}(r_p)$ averaged 
over $0.01~{\rm Mpc} \le r_p \le 10~{\rm Mpc}$}
\bigskip
\centerline{
\begin{tabular}{lcc}
\hline\hline
  & SDSS galaxies & MRS galaxies \\ \hline
Red           & $(3.74 \pm 1.61)\times 10^{-5}$ & $(2.97 \pm 1.69)\times 10^{-5}$ \\
Blue          & $(1.80 \pm 1.25)\times 10^{-5}$ & $(4.44 \pm 1.20)\times 10^{-5}$ \\
Bright        & $(4.88 \pm 1.35)\times 10^{-5}$ & $(4.64 \pm 2.07)\times 10^{-5}$ \\
Faint         & $(1.50 \pm 1.07)\times 10^{-5}$ & $(3.34 \pm 0.97)\times 10^{-5}$ \\
High $M_\ast$ & $(5.03 \pm 2.35)\times 10^{-5}$ & $(2.62 \pm 2.18)\times 10^{-5}$ \\
Low $M_\ast$  & $(1.93 \pm 1.69)\times 10^{-5}$ & $(3.61 \pm 0.98)\times 10^{-5}$ \\
High SSFR     & $(4.36 \pm 2.06)\times 10^{-5}$ & $(4.68 \pm 3.71)\times 10^{-5}$ \\
Low SSFR      & $(3.11 \pm 2.09)\times 10^{-5}$ & $(2.86 \pm 1.26)\times 10^{-5}$ \\ 
\hline
\end{tabular}
}

\end{document}